\documentclass[useAMS,usenatbib]{mn2e}

\usepackage[T1]{fontenc}
 \usepackage{pslatex}
\usepackage{amsmath}
\usepackage{amsfonts}
\usepackage{color}
\usepackage{url}
\usepackage{graphicx}
\usepackage{subfigure}
\usepackage{amssymb}
\usepackage{soul}
\usepackage{booktabs}
\usepackage{array}
\usepackage[utf8]{inputenc}
\inputencoding{latin1}
\inputencoding{utf8}

{\newif\ifnotend
\notendtrue
\def\veclist{ABCDEFGHIJKLMNOPQRSTUVWXYZabcdefghijklmnopqrstuvwxyz.}
\def\top#1#2.{#1}
\def\tail#1#2.{#2.}
\loop\expandafter\xdef\csname v\expandafter\top\veclist\endcsname%
{{\noexpand\bf\expandafter\top\veclist}}
\edef\veclist{\expandafter\tail\veclist}
\if\veclist.\notendfalse\fi\ifnotend\repeat}
\def\e{{\rm e}}
\def\i{{\rm i}}

\def\E{{\cal E}}

\mathchardef\mhyphen="2D

\title[Kink Instability of Force-Free Jets]{Kink Instability of Force-Free Jets: a Parameter Space Study}
\author[Sobacchi, Lyubarsky \& Sormani]{E. Sobacchi$^{1,2}$\thanks{E-mail: sobacchi@post.bgu.ac.il}, Y. E. Lyubarsky$^1$ \& M. C. Sormani$^3$\\
$^1$ Physics Department, Ben-Gurion University, P.O.B. 653, Beer-Sheva 84105, Israel \\
$^2$ Department of Natural Sciences, The Open University of Israel, 1 University Road, P.O.B. 808, Raanana 4353701, Israel\\
$^3$ Institute for Theoretical Astrophysics, Zentrum f\"{u}r Astronomie der Universit\"{a}t Heidelberg, Albert-\"{U}berle-Str. 2, 69120 Heidelberg, Germany
}
\begin{document}

\date{}

\def\p{\partial}
\def\E{\textbf{E}}
\def\B{\textbf{B}}
\def\v{\textbf{v}}
\def\j{\textbf{j}}
\def\s{\textbf{s}}
\def\e{\textbf{e}}

\newcommand{\di}{\mathrm{d}}
\newcommand{\bfx}{\mathbf{x}}
\newcommand{\bfe}{\mathbf{e}}
\newcommand{\vlos}{\mathrm{v}_{\rm los}}
\newcommand{\Tspin}{T_{\rm s}}
\newcommand{\Tb}{T_{\rm b}}
\newcommand{\degree}{\ensuremath{^\circ}}
\newcommand{\Th}{T_{\rm h}}
\newcommand{\Tc}{T_{\rm c}}
\newcommand{\bfr}{\mathbf{r}}
\newcommand{\bfv}{\mathbf{v}}
\newcommand{\bfu}{\mathbf{u}}
\newcommand{\pc}{\,{\rm pc}}
\newcommand{\kpc}{\,{\rm kpc}}
\newcommand{\Myr}{\,{\rm Myr}}
\newcommand{\Gyr}{\,{\rm Gyr}}
\newcommand{\kms}{\,{\rm km\, s^{-1}}}
\newcommand{\de}[2]{\frac{\partial #1}{\partial {#2}}}
\newcommand{\cs}{c_{\rm s}}
\newcommand{\rb}{r_{\rm b}}
\newcommand{\rqu}{r_{\rm q}}
\newcommand{\bfOmega}{\pmb{\Omega}}
\newcommand{\bfOmegap}{\pmb{\Omega}_{\rm p}}
\newcommand{\bfXi}{\boldsymbol{\Xi}}

\maketitle

\begin{abstract}
In the paradigm of magnetic acceleration of relativistic jets, one of the key points is identifying a viable mechanism to convert the Poynting flux into the kinetic energy of the plasma beyond equipartition. A promising candidate is the kink instability, which deforms the body of the jet through helical perturbations. Since the detailed structure of real jets is unknown, we explore a large family of cylindrical, force-free equilibria to get robust conclusions. We find that the growth rate of the instability depends primarily on two parameters: (i) the gradient of the poloidal magnetic field; (ii) the Lorentz factor of the perturbation, which is closely related to the velocity of the plasma. We provide a simple fitting formula for the growth rate of the instability.
As a tentative application, we use our results to interpret the dynamics of the jet in the nearby active galaxy M87. We show that the kink instability becomes non-linear at a distance from the central black hole comparable to where the jet stops accelerating. Hence (at least for this object), the kink instability of the jet is a good candidate to drive the transition from a Poynting-dominated to a kinetic-energy-dominated flow.
\end{abstract}



\section{Introduction}
\label{sec:introduction}

Astrophysical jets are ubiquitous in a wide variety of events, ranging from small-scale protostellar objects to large-scale extragalactic jets. The jets from microquasars (e.g. \citealt{MirabelRodriguez1999}), Active Galactic Nuclei (AGN; e.g. \citealt{UrryPadovani1995}) and Gamma Ray Bursts (GRBs; e.g. \citealt{Piran2004}) are accelerated to relativistic speeds.

One of the most promising explanations for jet launching is energy extraction from a rotating, magnetised source (e.g. \citealt{Blandford1976, Lovelace1976, BlandfordZnajek1977}). In the simplest scenario, magnetic field lines anchored to the central object act as sliding wires for the plasma that is accelerated by the magnetic tension. In this scenario, one of the key points is the fate of the magnetic fields at large distances from the source.
In the context of a steady, axisymmetric, ideal MHD flow, both analytical and numerical works have shown that the magnetic energy could be converted into the plasma kinetic energy up to equipartition (i.e. corresponding to a magnetisation $\sigma\sim 1$), but achieving further acceleration is generally difficult (e.g. \citealt{Komissarov2007, Komissarov2009, Lyubarski2009, Lyubarski2010, Lyubarski2011, Tchekhovskoy2008, Tchekhovskoy2009, Tchekhovskoy2010}).

Moreover, even for a relatively low magnetisation ($\sigma\gtrsim 0.1$), only weak shocks are possible, which makes the jet too radiatively inefficient to be consistent with observations of GRBs \citep{ZhangKobayashi2005, Mimica2009b, Mimica2009a, MimicaAloy2010, Narayan2011}. Spectral fitting of AGNs also require the plasma to be matter-dominated in the emission region, typically located at hundreds/thousands of gravitational radii from the black hole \citep{Ghisellini2010, Tavecchio2011}. Hence, it is crucial to identify some mechanism for efficient conversion of the magnetic energy into kinetic energy well beyond equipartition.

A possibility is that the instabilities in the MHD flow eventually destroy its regular structure and cause the release into the plasma of the energy stored in the magnetic fields (e.g. \citealt{Lyubarski1992, Eichler1993,Spruit1997,Begelman1998,Giannios2006}). In force-free jets, the most dangerous helical modes are indeed unstable if the poloidal magnetic field has a non-vanishing gradient \citep{IstominPariev1996, Lyubarski1999}.

In the context of relativistic jets, most of the analytical works focused on non-rotating flows (e.g. \citealt{Appl2000, Bodo2013}), or considered the limit of long/short wavelengths (e.g. \citealt{Lyubarski1999, Tomimatsu2001, Nalewajko2012}). \citet{Narayan2009} studied a two-parameter family of cylindrical, force-free equilibria, assuming a rigid impenetrable wall at the outer cylindrical radius; here we do not make this assumption, and we get complementary results.
For the more relevant case of rotating jets, in this paper we address (i) the dependence of the maximum growth rate of the instability on the gradient of the poloidal field (only the limit of long wavelengths has been studied so far); (ii) the relation between the group velocity of the perturbation and the velocity of the plasma. As a first step, we consider force-free jets.

In general, analytical results have proven extremely useful to interpret numerical simulations that explored the evolution of the kink instability in the non-linear regime and revealed its fundamental effect on the jet's structure after tens/hundreds of light crossing times (e.g. \citealt{Nakamura2007, Mizuno2009, Mizuno2012, Mizuno2014, Mignone2010, Oneill2012, Singh2016}).
Some consensus has eventually been reached on the fact that the kink instability can play an important role in the transition to a matter-dominated flow, at least for causally connected jets. Recently, different authors  \citep{PorthKomissarov2015, TchekhovskoyBromberg2016} even proposed that whether the jet becomes kink-unstable may explain the FRI/FRII dichotomy of radio galaxies \citep{FanaroffRiley1974}.

Since the detailed internal structure of real jets is unknown, it is important to identify the most fundamental physical parameters controlling how the instability develops. We attack this problem by selecting a large class of background solutions \citep{Mizuno2012}, and we explore a wide (three-dimensional) parameter space to get robust conclusions. We provide a simple analytic formula that approximately reproduces the growth rate of the kink instability and depends on (i) the gradient of the poloidal magnetic field; (ii) the Lorentz factor of the perturbation, which is closely related to the drift velocity of the plasma.
As a tentative application, we use our result to interpret the dynamics of the jet in the active galaxy M87, recently resolved down to hundreds of gravitational radii from the central black hole \citep{Mertens2016}.

The paper is organised as follows. In Section \ref{sec:equations} we present the background solution and the linearised equations for helical perturbations. In Section \ref{sec:results} we find the dispersion relation for the kink modes and we study in detail the group velocity and the growth rate of the instability. In Section \ref{sec:dis} we discuss how our results can be applied to the jet of the active galaxy M87. Finally, in Section \ref{sec:conclusions} we summarise our conclusions.

\section{Fundamental equations}
\label{sec:equations}

The fundamental equations governing relativistic jets in the ideal MHD approximation are the Maxwell's equations
\begin{align}
\label{eq:maxwell_1}
\nabla\times\E & = -\frac{1}{c}\frac{\p\B}{\p t} & \nabla\cdot\B & =0 \\
\label{eq:maxwell_2}
\nabla\cdot\E & =4\pi\rho& \nabla\times\B & =\frac{4\pi}{c}\j +\frac{1}{c}\frac{\p\E}{\p t}\;,
\end{align}
where $\rho$ and $\j$ are the charge and current densities. These are coupled with the condition of infinite conductivity
\begin{equation}
\label{eq:cond}
\E+\frac{\v}{c}\times\B=0 \;,
\end{equation}
where $\v$ is the velocity of the flow.
In the case of force-free flows, Euler's fluid equation reduces to
\begin{equation}
\label{eq:euler_forcefree}
\rho\E + \frac{\j}{c}\times\B=0\;.
\end{equation}

\subsection{Unperturbed solution}
\label{sec:background}

The equation for the steady-state equilibrium configuration of a cylindrical, force-free jet can be derived from Eq. \eqref{eq:maxwell_1} - \eqref{eq:euler_forcefree}. It is
\begin{equation}
\label{eq:equilibrium}
B_{\rm z}\frac{\text{d}B_{\rm z}}{\text{d}r}+\frac{B_\phi}{r}\frac{\text{d}}{\text{d}r}\left(rB_\phi\right)-\frac{\Omega B_{\rm z}}{c}\frac{\text{d}}{\text{d}r}\left(\frac{\Omega r^2B_{\rm z}}{c}\right) =  0 \;,
\end{equation}
where 
\begin{equation}
\label{eq:defomega}
\Omega \equiv \frac{1}{r}\left(v_\phi - \frac{B_\phi}{B_{\rm z}} v_{\rm z}\right)
\end{equation}
is the angular velocity of the field lines (with this definition $E_{\rm r}~=~\Omega rB_{\rm z}/c$). Note that, since the toroidal magnetic field is generated by the rotation of the base of the jet, the sign of $B_\phi$ is opposite to the sign of $\Omega B_{\rm z}$ (for jets propagating in the positive direction).

We consider a poloidal magnetic field of the form
\begin{equation}
\label{eq:Bz}
B_{\rm z}=\frac{B_0}{\left[1+\left(r/r_0\right)^2\right]^\alpha}\;,
\end{equation}
where $r_0$ is the typical scale of the jet core and the parameter $\alpha$ defines the poloidal field profile (e.g. \citealt{Mizuno2012}). Note that $\alpha$ is also equal to minus the logarithmic derivative of $B_{\rm z}$ calculated at the core scale, $r=r_0$.

With this poloidal field, Eq. \eqref{eq:equilibrium} has an analytical solution for the toroidal field
\begin{equation}
\label{eq:Bphi}
B_\phi = \frac{B_0}{\left[1+\left(r/r_0\right)^2\right]^\alpha}\times\left[P^2 + \left(\frac{\Omega r}{c}\right)^2\right]^{1/2}\;,
\end{equation}
where
\begin{equation}
\label{eq:pitch}
P\equiv\left[\frac{\left(r_0/r\right)^2\left[1+\left(r/r_0\right)^2\right]^{2\alpha}-\left(r_0/r\right)^2-2\alpha}{2\alpha-1}\right]^{1/2}
\end{equation}
is the ratio between the toroidal and poloidal components of the magnetic field in a non rotating jet.
In the following we parametrize the angular velocity as
\begin{equation}
\label{eq:omega}
\Omega=-\frac{\Omega_0}{1+\left(r/r_0\right)^\beta} \;,
\end{equation}
where $\beta$ defines its slope at large radii.

Both the current density and the Poynting flux corresponding to Eq. \eqref{eq:Bz} - \eqref{eq:omega} are peaked at $r\sim r_0$. Nevertheless, this solution may formally give a finite/diverging total current, depending on the asymptotic (i.e. $r\gg r_0$) profile of the fields. The current flowing through a circular section of the jet with radius $r$ is $I=rB_\phi/2c$. Since from Eq. \eqref{eq:Bphi} one can show that $B_\phi$ is never decaying faster than $1/r$, the total current cannot formally vanish at infinity.
However, if the fields sharply decline outside of some radius $r_1\gg r_0$ (which we may associate with the typical size of the accretion disk, whose magnetic field is confining the jet), the total current vanishes. Such a scenario does not significantly change the fields/currents in the most relevant region within a few $r_0$, but produces small currents at large radii (i.e. $r\sim r_1$) which balance the total flux. Since the perturbations are well localised at the core, this slight modification does not affect their time evolution.

\subsection{Linearised equations for the perturbed flow}

Stability of cylindrical equilibria can be investigated considering perturbations on the magnetic surfaces of the form
\begin{equation}
\label{eq:pert}
f\left(r\right)\exp\left[\i\left(\omega t + m\phi - kz\right)\right]\;.
\end{equation}
In this paper we use the method developed by \citet{Solovev1967} for non-relativistic MHD flows. \citet{Lyubarski1999} has adapted this method to force-free jets, showing that the time evolution of the perturbation is described by a second order linear differential equation, namely
\begin{equation}
\label{eq:main}
\frac{\text{d}}{\text{d}r}\left[G\frac{\text{d}f}{\text{d}r}\right]=Df\;.
\end{equation}
The functions $G$ and $D$ can be expressed in terms of the unperturbed solution as
\begin{align}
\label{eq:G}
G & \equiv \frac{r^3}{1-u^2}\left(\frac{a}{\beta}-b\right) \\
\label{eq:D}
D & \equiv k^2\left[\left(1-u^2+\frac{m^2-1}{k^2r^2}\right)G - d - \frac{1}{\beta}\frac{\text{d}}{\text{d}r}\left(r^4b\right)\right] \;,
\end{align}
where
\begin{align}
a & \equiv \left[\left(1-muV-u^2\right)B_{\rm z}-\frac{m}{kr}B_\phi\right]^2 \\
b & \equiv \left[VB_{\rm z}+\frac{u}{kr}B_\phi\right]^2 \\
d & \equiv \frac{2r^3}{\beta^2} \left[ \left(muVB_{\rm z}+\frac{m}{kr}B_\phi\right)^2  -  \left(1-u^2\right)^2B_{\rm z}^2\right] \;.
\end{align}
Here we have defined $V\equiv\Omega/kc$ and $u\equiv\omega/kc$. In general, the solution of Eq. \eqref{eq:main} needs to satisfy two homogeneous boundary conditions, namely (i) an arbitrary normalisation; (ii) $f$ vanishing at infinity. Hence, the value of $\omega$ is automatically determined by these requirements.

In the following we focus on the $m=1$ mode, corresponding to the fastest growing instability (e.g. \citealt{Bateman1978}). For completeness, we include a brief discussion of modes with higher $m$ in Appendix \ref{sec:appendixA}.
When $u$ is complex (the relevant case since we aim to study jet instabilities) the function $G$ does not vanish at any radius $r>0$. For $m=1$, at $r\sim 0$ we have $G \sim D \sim r^3$; therefore, expansion at the first order close to the origin implies $f'\left(0\right)=0$. With an arbitrary normalisation (e.g. $f\left(0\right)=1$) we are left with the initial conditions of a standard Cauchy problem, and Eq. \eqref{eq:main} can then be integrated numerically to find the solution for any $r>0$.\footnote{Numerical integration is started from a positive $r_{\rm init}\ll r_0$ (where $r_0$ is the typical scale of the jet core). Taylor expansion of the solution is used to find the proper initial conditions.}

However, the asymptotic behaviour of the solution poses an additional constraint. For a given mode, when $r\gg 1/k$ we have
$D\sim k^2\left(1-u^2\right)G$. Hence, in this limit $f\sim\exp\left(\pm k\sqrt{1-u^2} r\right)$. Since we are looking at small perturbations (i.e. $f\ll r$), the physical solution is the decaying exponential and we need the outer boundary condition $\lim_{r\to\infty} f\left(r\right)=0$.\footnote{We use a finite $r_{\rm end}\gg r_0$ for the outer boundary condition, namely $f\left(r_{\rm end}\right)=0$. We have checked that the choice of $r_{\rm init}$ and $r_{\rm end}$ does not affect the results.}
The value of $u$ matching this further condition can be found with the standard shooting method for eigenvalue problems (e.g. \citealt{Press2002}).

\section{Results}
\label{sec:results}

\begin{figure}{\vspace{3mm}} 
\centering
\includegraphics[width=0.49\textwidth]{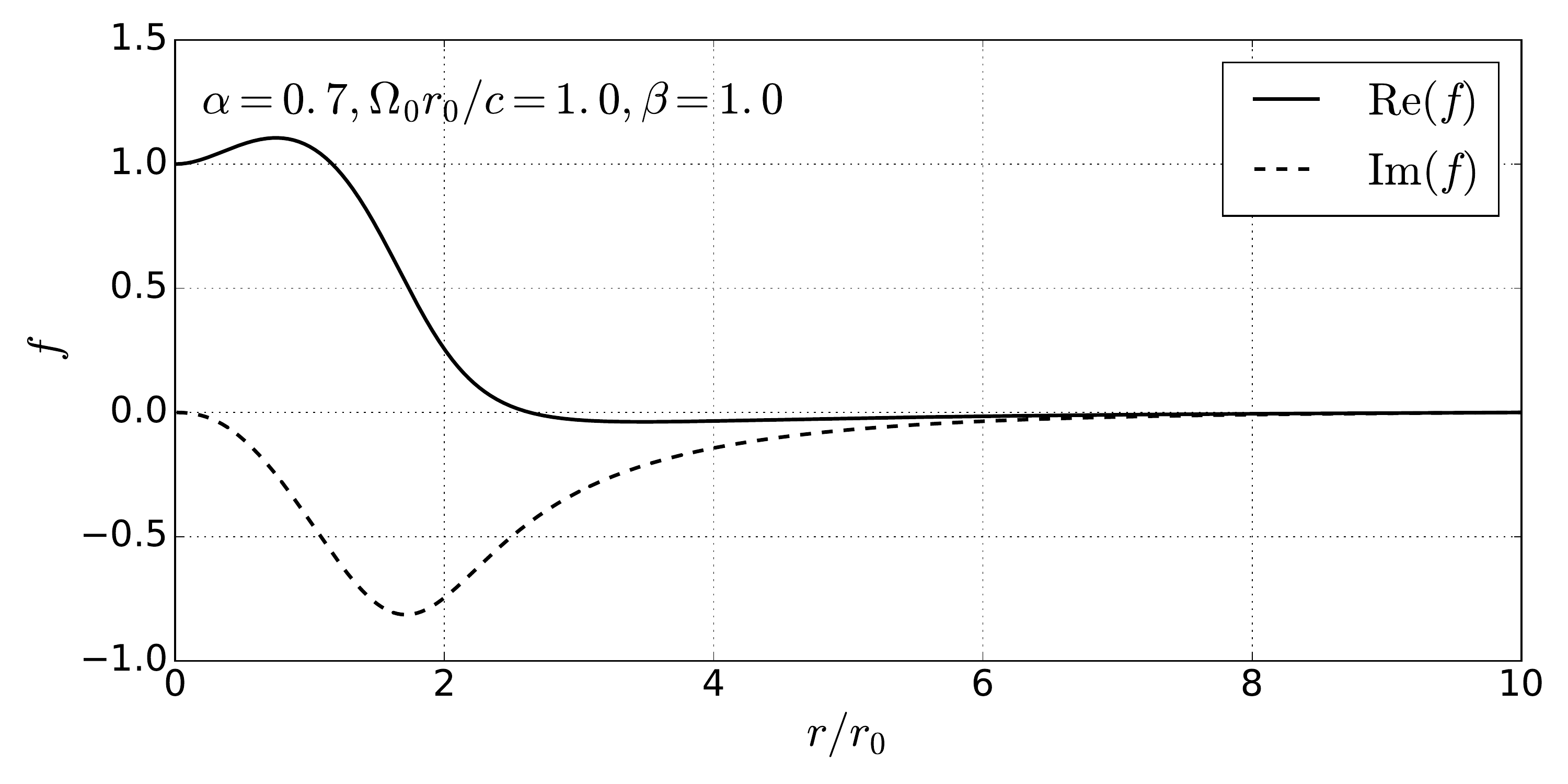}
\caption{Solution of Eq. \eqref{eq:main} for our fiducial parameters $\alpha=0.7$, $\Omega_0r_0/c~=~1.0$, $\beta=1.0$; the wavelength is given by $kr_0=0.7$. Solid/dashed lines correspond to the real/imaginary parts of $f$ respectively.}
\label{fig:shooting}
\end{figure}

This section is dedicated to study the solution of Eq. \eqref{eq:main} for different configurations of the background fields. In Figure \ref{fig:shooting} we show the solution of Eq. \eqref{eq:main} when $\alpha=0.7$, $\Omega_0r_0/c=1.0$, $\beta=1.0$ (solid/dashed lines correspond to the real/imaginary parts of $f$ respectively). The wavelength is comparable to the size of the core (we use $kr_0=0.7$). Note that the perturbation is well concentrated at the core of the jet, and $f$ practically vanishes at $r\sim 10\;r_0$.

\subsection{Dispersion relation}

\begin{figure*}{\vspace{3mm}} 
\centering
\includegraphics[width=0.49\textwidth]{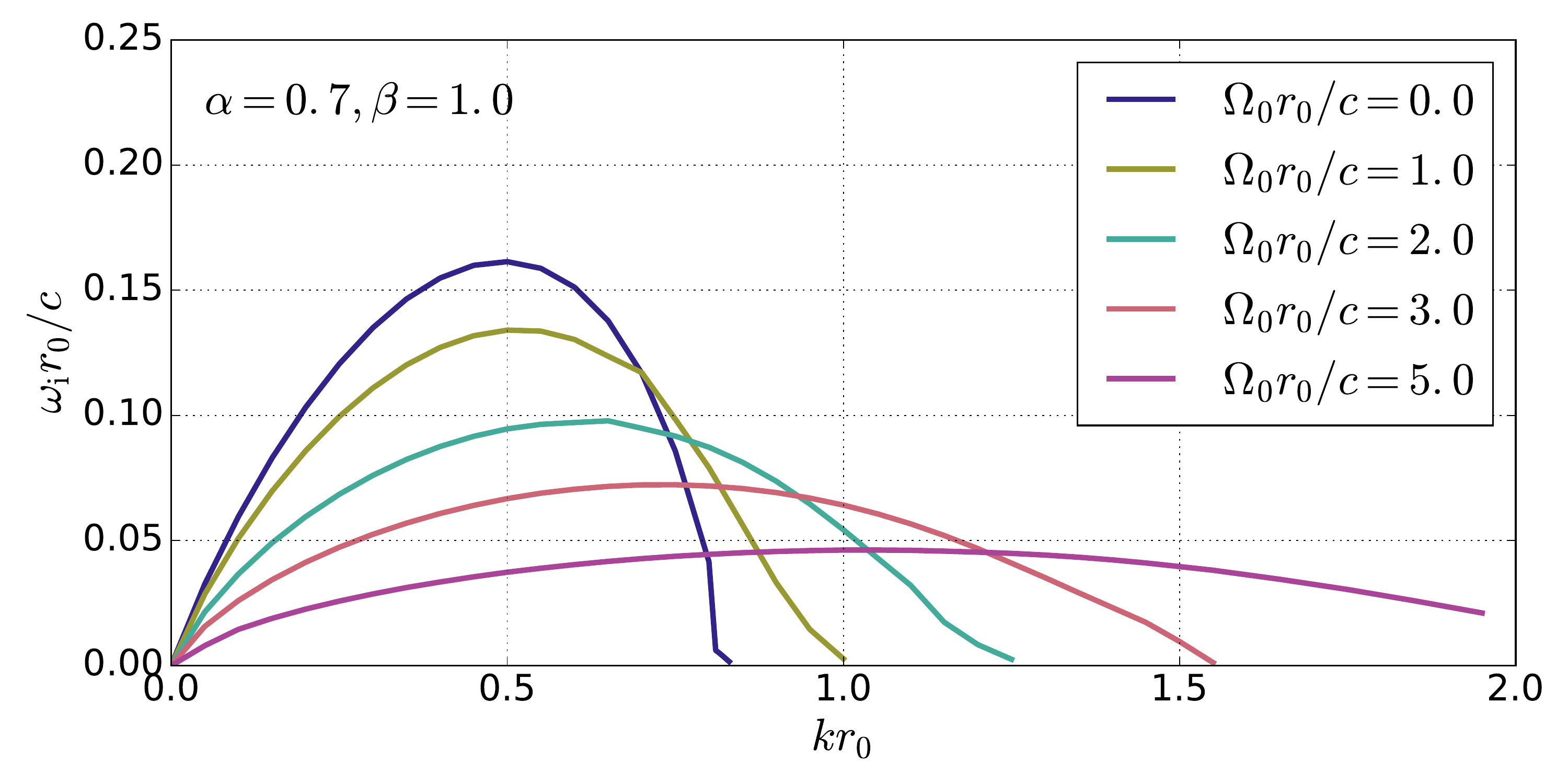}
\includegraphics[width=0.49\textwidth]{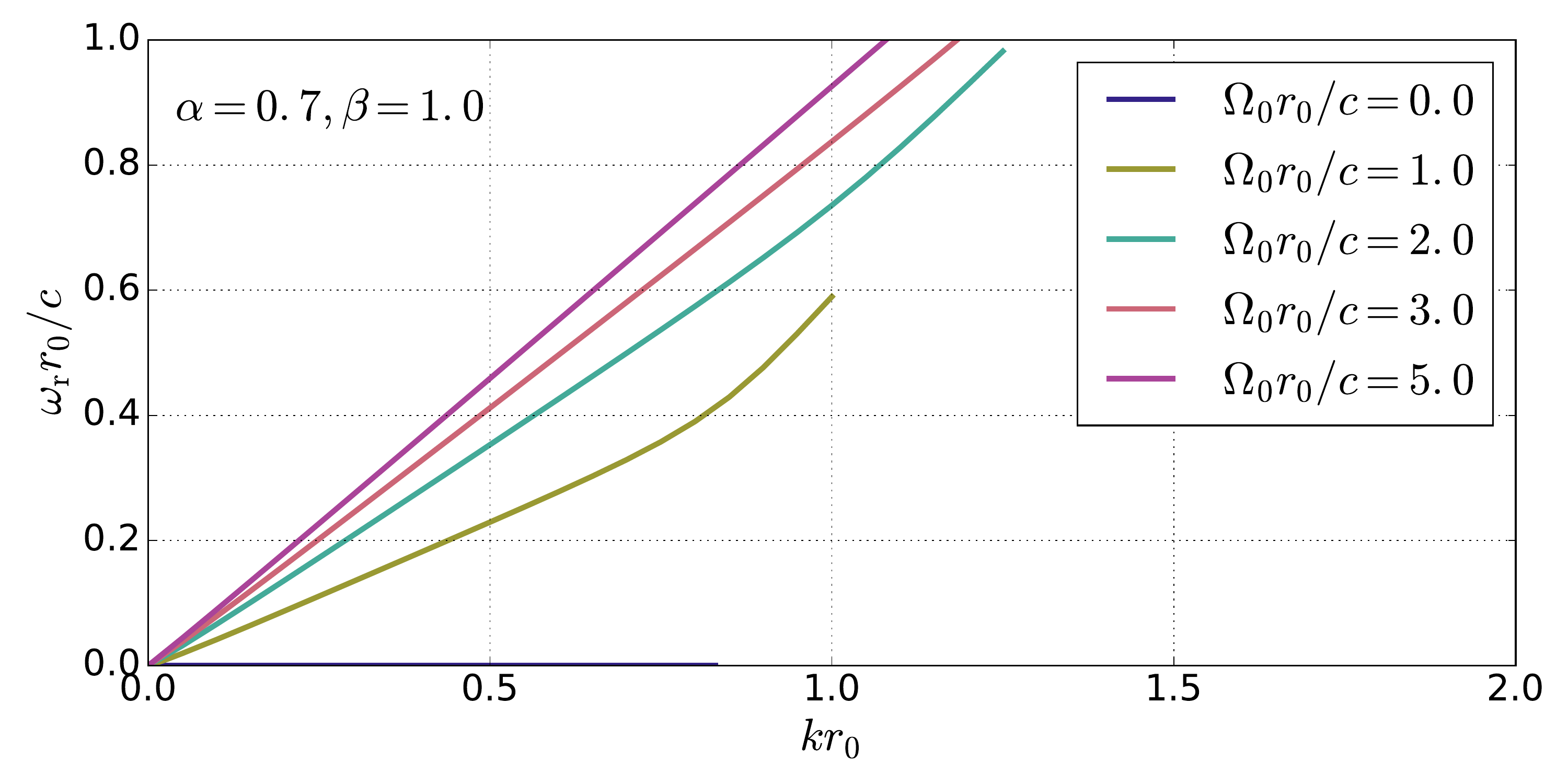}
\includegraphics[width=0.49\textwidth]{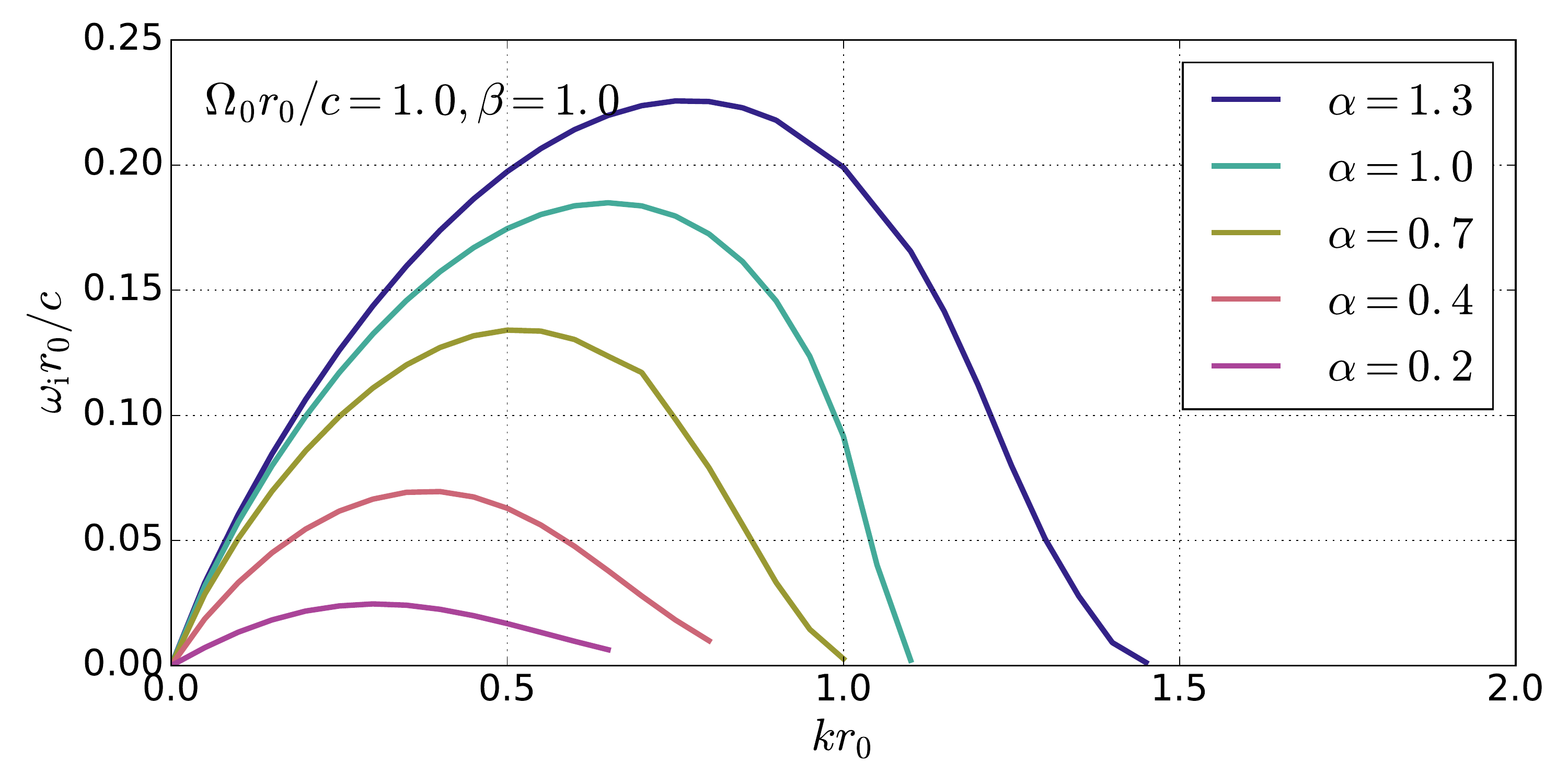}
\includegraphics[width=0.49\textwidth]{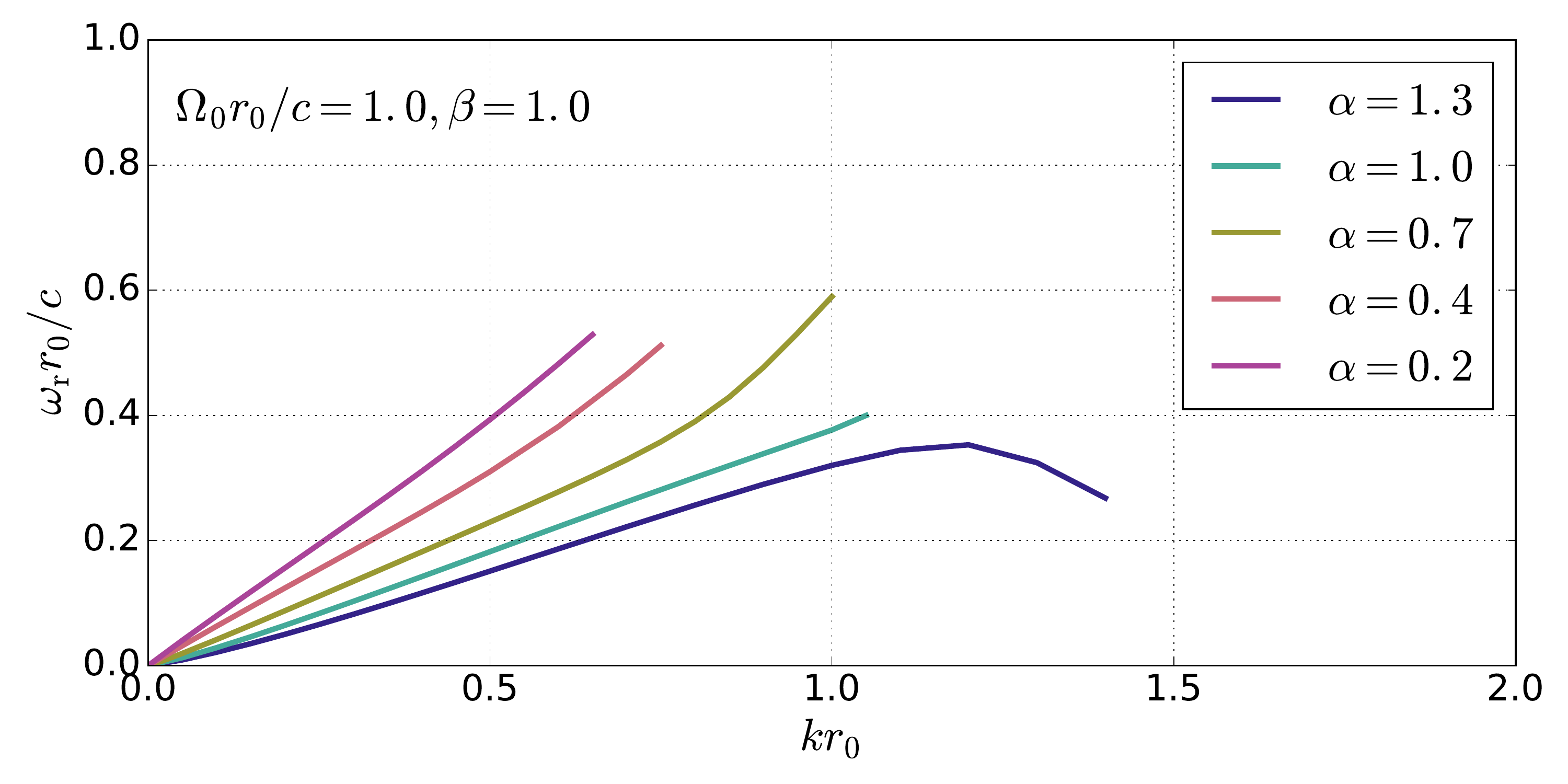}
\includegraphics[width=0.49\textwidth]{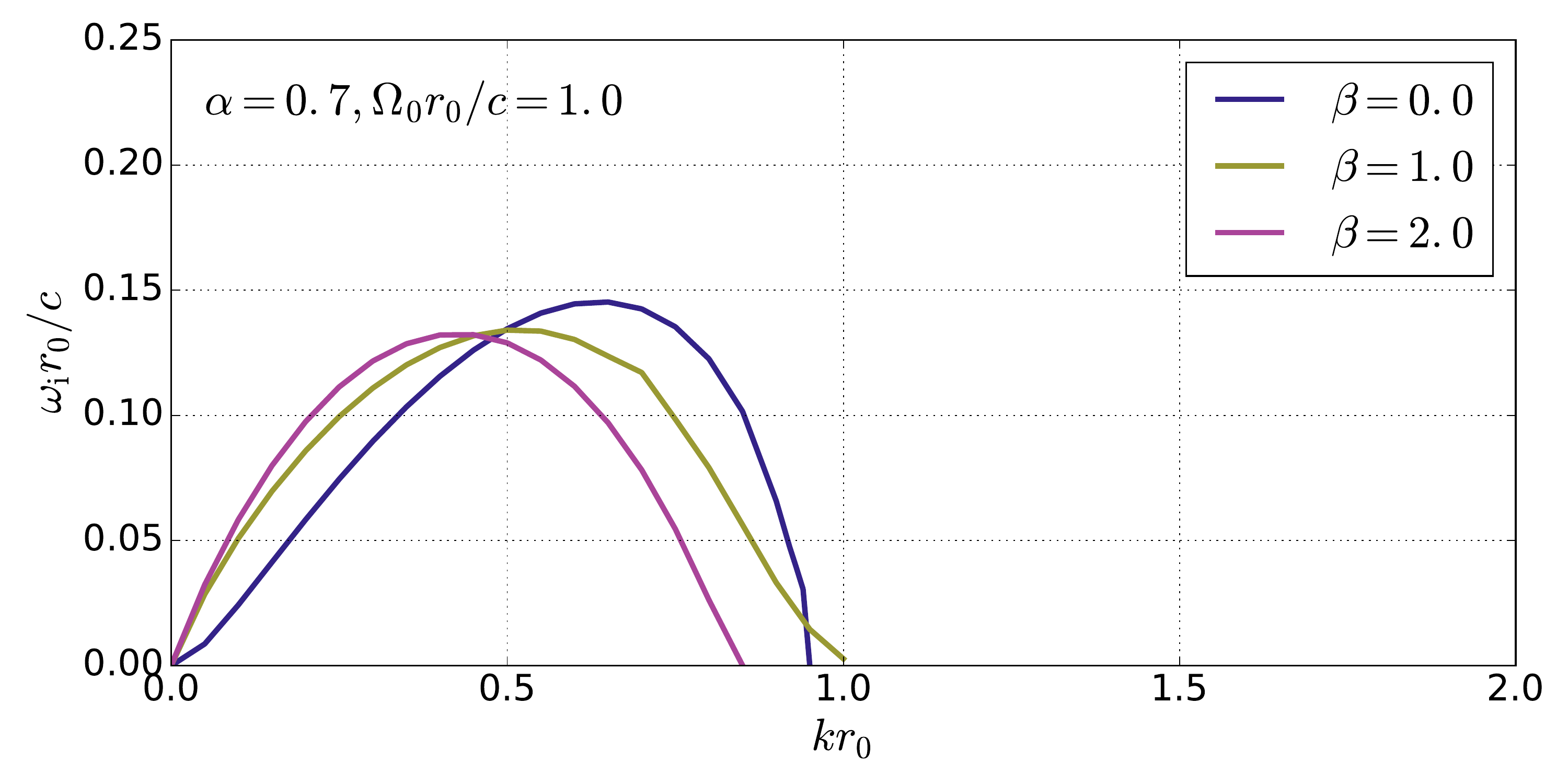}
\includegraphics[width=0.49\textwidth]{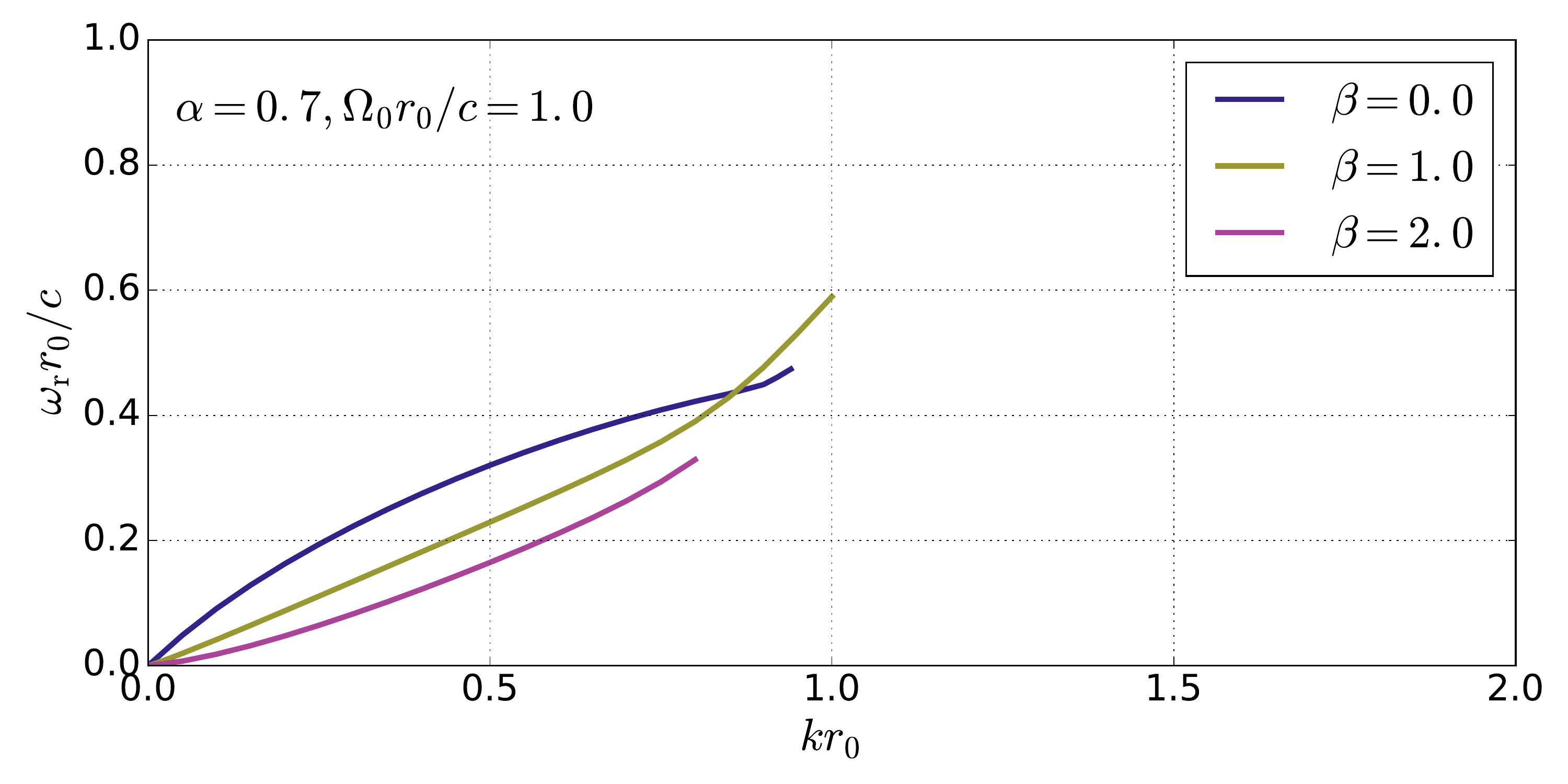}
\caption{Dispersion relation for the imaginary (left) and real (right) part of $\omega$ in a force-free jet. From top to bottom we show the effect of changing the angular velocity ($\Omega_0$); the magnetic field profile parameter ($\alpha$); the asymptotic slope of the angular velocity ($\beta$), while the other two parameters are fixed. As a fiducial model, we take $\Omega_0 r_0/c=1.0$, $\alpha=0.7$ and $\beta=1.0$.}
\label{fig:dispersion}
\end{figure*}

In Figure \ref{fig:dispersion} we show the dispersion relation in a force-free jet for the imaginary and real parts of $\omega$ ($\omega_{\rm i}$/$\omega_{\rm r}$ in left/right panels respectively). As a fiducial model, we take $\Omega_0 r_0/c=1.0$, $\alpha=0.7$ and $\beta=1.0$. Then, we keep two of these parameters fixed and we study the effect of changing the other one ($\Omega_0$, $\alpha$ and $\beta$ from top to bottom).

First of all note that the most unstable wavelengths are comparable but longer than the scale of the jet core ($kr_0\sim 0.3-0.8$), and the typical growth rate of the instability is some fraction of the light crossing frequency ($\omega_{\rm i}r_0/c \lesssim 0.25$ for the parameter range considered here).

For a given triplet $\left(\Omega_0;\alpha;\beta\right)$ the right panels in Figure \ref{fig:dispersion} show that the group velocity of the perturbation is nearly constant over almost the entire unstable region. However, it shows significant variations at the shortest unstable wavelengths (largest $k$), also exceeding $c$ by $\lesssim 10\%$ in few cases. This is due to the fact that, when
$\omega_{\rm i}r_0/c \ll 1$, the solution of Eq. \eqref{eq:main} has a resonance which is difficult to treat numerically. Since these complications arise only in a limited $k$ region (where the instability is weak), they do not affect the general behaviour of the perturbation.

The maximum growth rate of the instability, $\omega_{\rm i,max}$, decreases with $\Omega_0$, while the group velocity of the perturbation, $v_{\rm p}\equiv \text{d}\omega_{\rm r}/\text{d}k$ (calculated for the most unstable $k$), increases. For example, $\omega_{\rm i,max}r_0/c=0.16-0.046$ and $v_{\rm p}/c=0-0.94$ when $\Omega_0 r_0/c=0-5$. This result is consistent with the analytical study of \citet{Lyubarski1999}, who found a relativistic suppression of the instability in the limit of long wavelengths. Note that the most unstable $k$ increases with $\Omega_0$ in the same range.

When the poloidal field is nearly flat (i.e. small $\alpha$), the instability is growing slowly, and completely disappears when $\alpha=0$ \citep{IstominPariev1996, Lyubarski1999}. The poloidal field profile parameter has a strong impact on the dispersion relation: there is one order of magnitude difference in the growth rate ($\omega_{\rm i,max}r_0/c=0.023-0.25$) when $\alpha=0.2-1.3$. In the same range, the group velocity is decreasing, though less significantly ($v_{\rm p}=0.77-0.36$).

Both the maximum growth rate and the group velocity of the instability are almost independent on the asymptotic slope of the angular velocity, $\beta$. In the range $\beta=0-2$, the variation is $\sim10\%$ for $\omega_{\rm i,max}$ and $\sim 25\%$ for $v_{\rm p}$ (note that the difference in $\omega_{\rm r}$ is larger, but the slope for the most unstable $k$ is almost constant).

\subsection{Group velocity of the perturbation}

The Lorentz factor of the perturbation (calculated for the most unstable wavelength, $\gamma_{\rm p}\equiv1/\sqrt{1-v_{\rm p}^2/c^2}$) is fundamental to understand its time evolution. Therefore, one would like to connect it to the Lorentz factor of the plasma, which is easier to visualise physically. The problem is that in force-free jets the Lorentz factor is generally undetermined.

However, sufficiently far from the source the velocity of the plasma approaches a pure drift motion in the electromagnetic fields, i.e. $\textbf{v}/c\simeq\textbf{E}\times\textbf{B}/B^2$ \citep{Tchekhovskoy2009}. In this case, for our setup one can easily calculate the Lorentz factor of the plasma as
\begin{equation}
\label{eq:gamma}
\gamma_{\rm drift}=\left[1+\frac{1}{1+P^2}\left(\frac{\Omega r}{c}\right)^2\right]^{1/2} \;,
\end{equation}
where $P$ and $\Omega$ are given by Eq. \eqref{eq:pitch} and \eqref{eq:omega} respectively. Since
$\gamma_{\rm drift}$ depends on the radius, we take the Lorentz factor at the core scale, $\gamma_{\rm drift,0}\equiv\gamma_{\rm drift}\left(r_0\right)$, as a proxy for the typical Lorentz factor of the plasma. It is possible to show that $\gamma_{\rm drift}$ peaks around this scale, and that the drift velocity eventually vanishes at infinity (i.e. when $r\gg r_0$).

To study the connection between $\gamma_{\rm p}$ and $\gamma_{\rm drift,0}$, we select $15$ random points in our three-dimensional parameter space (in the range $0<\Omega_0r_0/c<10$; $0.2<\alpha<1.3$; $0<\beta<2$) and we calculate the corresponding dispersion relations. The upper limit on $\Omega_0$ and the lower limit on $\alpha$ are selected in order to avoid the case $\omega_{\rm i,max}r_0/c\ll 1$, which is difficult to treat numerically.

\begin{figure}{\vspace{3mm}} 
\centering
\includegraphics[width=0.49\textwidth]{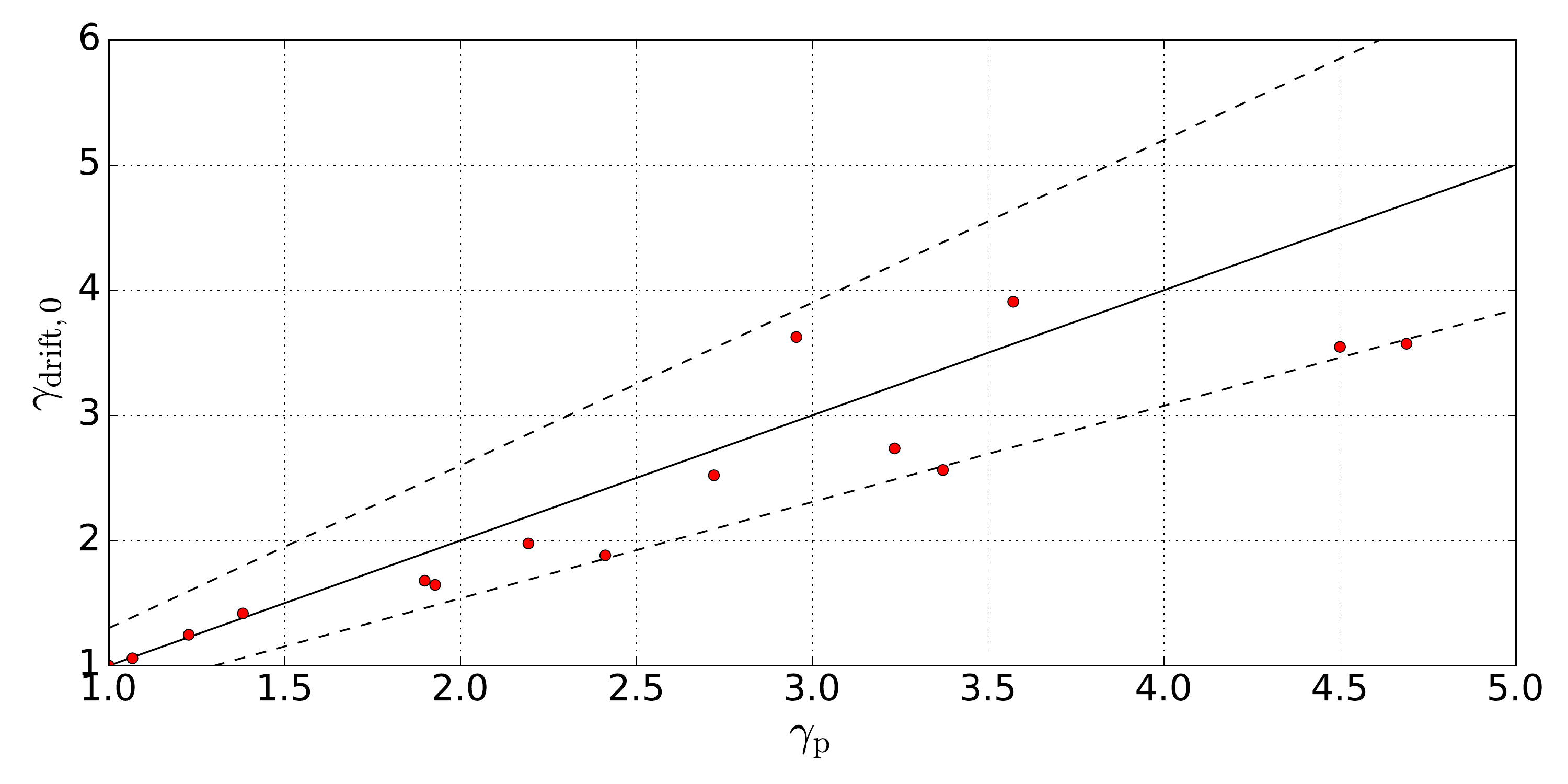}
\caption{Correlation between the Lorentz factor of the perturbation ($\gamma_{\rm p}$, calculated for the most unstable wavelength) and of the plasma flow for a pure drift motion ($\gamma_{\rm drift}$); since the velocity of the plasma depends on the radius, $\gamma_{\rm drift}$ is calculated at the core scale, i.e. $\gamma_{\rm drift,0}\equiv\gamma_{\rm drift}\left(r_0\right)$.
The relevant parameters for different points are chosen randomly in the range $0<\Omega_0r_0/c<10$; $0.2<\alpha<1.3$; $0<\beta<2$.
The solid (dashed) lines, corresponding to $\gamma_{\rm drift,0}=\gamma_{\rm p}$ ($\gamma_{\rm drift,0}=\gamma_{\rm p}/1.3$ and $\gamma_{\rm drift,0}=1.3\times\gamma_{\rm p}$), help visualisation.}
\label{fig:main_gamma}
\end{figure}

In Figure \ref{fig:main_gamma} we plot the Lorentz factor of the perturbation ($\gamma_{\rm p}$) versus the typical Lorentz factor of a pure drift motion ($\gamma_{\rm drift,0}$) for the 15 different choices of the parameters.
As we can see, $\gamma_{\rm p}$ and $\gamma_{\rm drift,0}$ are clearly correlated; this is a natural result since the instability develops in the plasma comoving frame. In particular, for all the combinations of parameters we used, we found $\gamma_{\rm drift,0}/1.3<\gamma_{\rm p}<1.3\times\gamma_{\rm drift,0}$ (dashed lines in the figure). Hence, one can identify the two Lorentz factors within reasonable accuracy.

\subsection{Growth rate of the instability}

\begin{figure}{\vspace{3mm}} 
\centering
\includegraphics[width=0.49\textwidth]{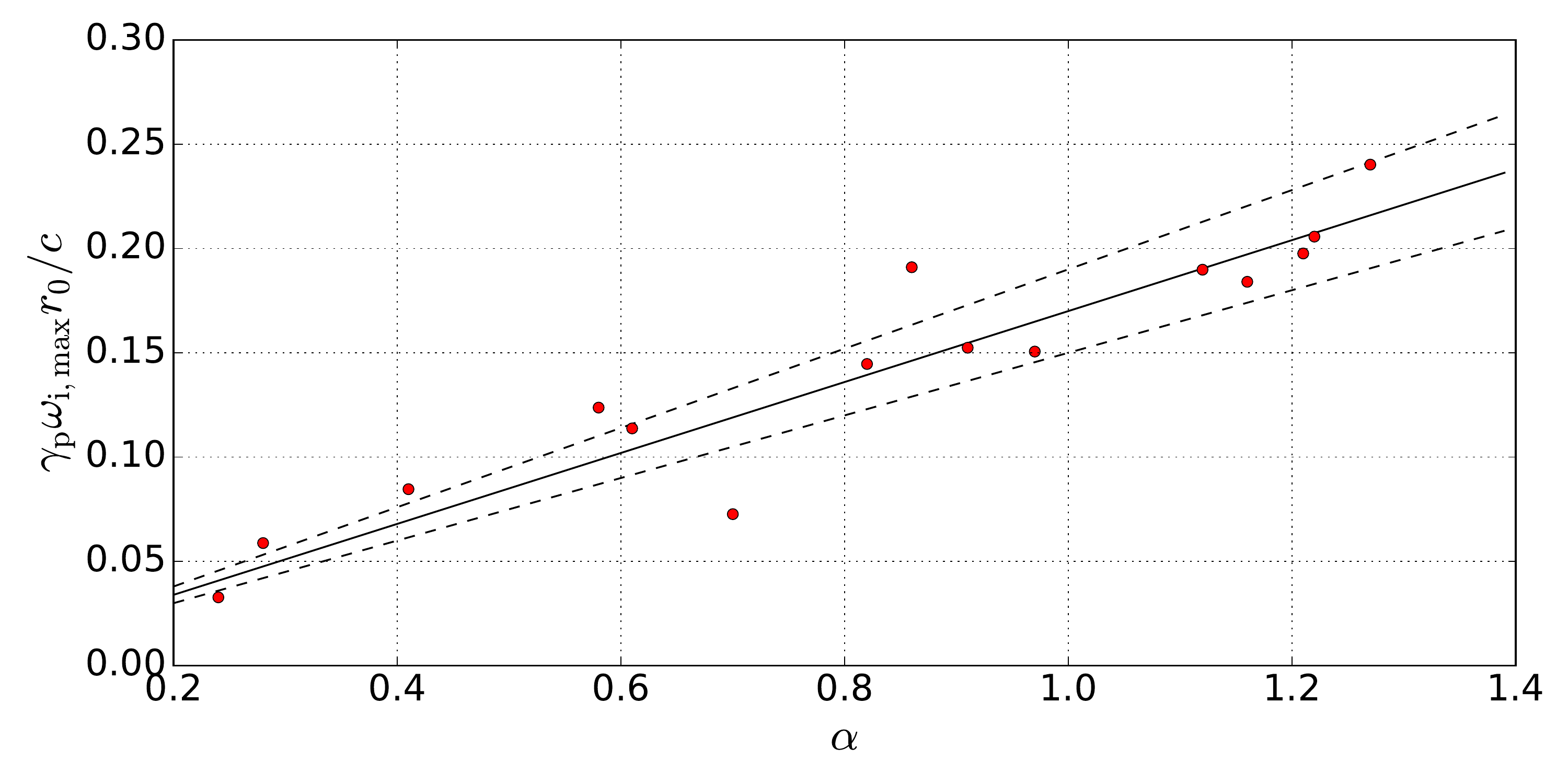}
\caption{Dependence of the maximum growth rate of the instability (rescaled for relativistic suppression) on the poloidal field profile parameter $\alpha$. The relevant parameters for different points are the same as in Figure \ref{fig:main_gamma}. The lines show the approximation from Eq. \eqref{eq:fit} with different coefficients (see text for details).}
\label{fig:main_alpha}
\end{figure}

In general, it would be useful to have a quick way to estimate the growth rate of the instability, without the need to recalculate the dispersion relation for each different jet structure we are interested to study. This is particularly true while considering realistic jets, where the detailed structure of the fields is unknown and we necessarily rely on order-of-magnitude estimates of the relevant physical parameters.

In Figure \ref{fig:main_alpha}, the maximum growth rates (in the rest frame of the perturbation, $\gamma_{\rm p}\omega_{\rm i,max}r_0/c$) for the 15 background solutions considered above are shown as dots as a function of $\alpha$.
Since there is a relatively small scatter at a given $\alpha$, it is possible to estimate the growth rate of the instability as
\begin{equation}
\label{eq:fit}
\frac{\omega_{\rm i,max}r_0}{c}=\frac{f\left(\alpha\right)}{\gamma_{\rm p}} \;,
\end{equation}
where the solid line in Figure \ref{fig:main_alpha} corresponds to
\begin{equation}
\label{eq:fit_aux}
f\left(\alpha\right)= 0.17\times\alpha \;.
\end{equation}
For comparison, the dashed lines show the same Eq. \eqref{eq:fit_aux} with different coefficients ($0.15$ and $0.19$ for the lower/upper curves). Note that the growth rate vanishes when $\alpha=0$ (i.e. for a flat poloidal field).
This is particularly important while interpreting the results of numerical simulations, since the stability of the jet crucially depends on the gradient of the poloidal field at the core.

Using Eq. \eqref{eq:fit} we can calculate the characteristic growth time of the instability as
\begin{equation}
\label{eq:t_growth}
T_{\rm i}\equiv\frac{1}{\omega_{\rm i,max}} \approx 25\times\left(\frac{\gamma_{\rm p}}{3}\right)\left(\frac{0.7}{\alpha}\right)\frac{r_0}{c}\;,
\end{equation}
where we have substituted $f\left(\alpha\right)$ from Eq. \eqref{eq:fit_aux}. In general, one would expect the kink instability to become non-linear after few $T_{\rm i}$. This result is in general agreement with numerical simulations with a similar setup, typically finding that the kink instability significantly develops for a time $\approx 100\times r_0/c$, before saturating in the fully non-linear regime (e.g. \citealt{Mizuno2009, Mizuno2012, Mizuno2014, Oneill2012, Singh2016}).
Moreover, \citet{Mizuno2012} explicitly considered the effect of the poloidal field gradient on the perturbation, showing that for small $\alpha$ the instability is severely suppressed also in the non-linear regime.

\section{Implications for the jet of M87}
\label{sec:dis}

In the paradigm of magnetic launching, the kink instability is often invoked to explain how a jet can transfer its energy from the Poynting flux to the kinetic energy of the plasma. For this mechanism to be efficient, the jet needs to be strongly causally connected, i.e.
$\theta_{\rm jet}\Gamma\lesssim 1$, where $\theta_{\rm jet}$ and $\Gamma$ are the opening angle and the Lorentz factor of the jet (e.g. \citealt{Komissarov2009, Lyubarski2009, Tchekhovskoy2009, Granot2011, PorthKomissarov2015}). This condition can be verified in AGN (though with a large scatter; e.g. \citealt{Pushkarev2009, Clausen2013}), while it is violated by GRBs (\citealt{KumarZhang2015} and references therein).\footnote{\citet{MckinneyBlanford2009} performed a 3D simulation of a rapidly rotating black hole producing an approximately conical jet, with $\theta_{\rm jet}\sim 5\degree$ and $\Gamma\sim 10$. They found this jet to be stable, retaining a high magnetisation out to $10^3$ gravitational radii despite small wiggles interpreted as a signature of the kink modes. Interestingly, this jet is at the boundary for strong causality (they have $\theta_{\rm jet}\Gamma\sim 0.9$). Moreover, the kink instability typically becomes non-linear around the largest scale they simulated (see below).}

A fundamental prototype of collimated jet is that in the active galaxy M87. \citet{Mertens2016} recently resolved the dynamics of this jet down to hundreds of gravitational radii from the central black hole. This galaxy is therefore an ideal case to study the mechanisms acting during the launching of the jet. The jet has an approximately parabolic shape, and the plasma accelerates linearly out to $L_{\rm f} \sim 1000\; r_{\rm g}$ (where $r_{\rm g}$ is the gravitational radius) from the source. For the central black hole we assume a mass of $\sim 3.5\times 10^9M_\odot$ (e.g. \citealt{Walsh2013}), corresponding to a gravitational radius $r_{\rm g}\sim 10^{15}\text{ cm}$. During this phase, the typical transverse scale of the jet spans a range $r_0\sim 30-90\; r_{\rm g}$ (see their Figure 6). After the linear acceleration phase, the Lorentz factor slowly increases for several orders of magnitude in distance.

Even in perfectly collimated jets, the kink instability becomes non-linear only at a distance $L_{\rm i}\sim cT_{\rm i}$ from the source. Using  $r_0~\sim~30-90\; r_{\rm g}$, our Eq. \eqref{eq:t_growth} gives
\begin{equation}
\label{eq:l_growth}
L_{\rm i}\sim cT_{\rm i}\approx 700-2000\times \left(\frac{\Gamma}{3}\right) \left(\frac{0.7}{\alpha}\right) r_{\rm g} \;,
\end{equation}
where, according to the discussion above, we have identified the Lorentz factors of the perturbation and of the plasma.\footnote{Formally, our Eq. \eqref{eq:t_growth} was derived considering force-free jets. Of course, the effect of a finite magnetisation (even if $\sigma\gtrsim 1$) and the presence of a confining external medium can affect the result. However, we believe that at least the order of magnitude is preserved.} Interestingly, this length scale is comparable with the end of the linear acceleration regime, i.e. 
\begin{equation}
L_{\rm i}\approx L_{\rm f}\;.
\end{equation}

This suggests the following scenario: (i) close to the central engine, the flow accelerates while dominated by the Poynting flux, or at most in a state of equipartition; (ii) at a typical distance $\sim 1000\; r_{\rm g}$ the kink instability enters its non-linear regime, eventually transferring the energy of the jet from the Poynting flux to the plasma (note that also blazar observations require the energy conversion to be completed around this scale; e.g. \citealt{Ghisellini2010, Tavecchio2011}); (iii) farther away, the jet is dominated by the kinetic energy of the plasma and the acceleration almost stops.

Of course, additional (potentially large) uncertainties are due to the unknown value of $\alpha$. However, at least when the jet is accelerated from $\sigma\gg 1$ down to $\sigma\approx 1$, the poloidal flux is concentrated in the vicinity of the axis \citep{BeskinNokhrina2009, Lyubarski2009}, and a typical $\alpha\approx 1$ seems a reasonable description for the core of the jet \citep{Tchekhovskoy2009}.

In general, one would expect a strong dissipation in the region where the global structure of the jet is destroyed by the kink modes. Unfortunately, resolving the jet down to these scales in wavebands different than the radio is not feasible with current facilities. However, the time variability of the light curves provides interesting constraints on the size of the emitting region. For M87, variabilities on extremely short time scales ($t_{\rm var}\approx 2\text{ days}$) have been detected in the TeV region, a factor $\sim 10$ faster than in other bands \citep{Aharonian2006}. This puts an upper limit on the size of the emitting region, $r_{\rm em}\lesssim \delta ct_{\rm var}$ (where $\delta\equiv 1/\left[\Gamma\left(1-\beta\cos\theta_{\rm obs}\right)\right]$ is the Doppler factor of the jet and $\Gamma\equiv1/\sqrt{1-\beta^2}$). Using $\Gamma\sim 3$ and a viewing angle $\theta_{\rm obs}\sim 17\degree$ \citep{Mertens2016},\footnote{Note that for M87 we have $\theta_{\rm obs}\lesssim 1/\Gamma$. Hence, it is possible to observe the emitted radiation despite beaming.} one eventually finds $r_{\rm em}\lesssim 40\; r_{\rm g}$.
Since this upper limit is comparable with the typical transverse scale of the jet, $r_0\sim 30-90\; r_{\rm g}$, the bulk of the emission may come from the same region where the kink instability becomes non-linear.

In the first regime of the outlined scenario (i.e. Poynting-dominated flow), theoretical models predict oscillations of the jet cross section \citep{Lyubarski2009}; alternatively, some oscillations may be due to the kink modes while still in the linear regime. Also in the third regime, the Kelvin-Helmholtz instability of the kinetic-energy-dominated plasma can result in similar patterns (\citealt{Hardee2000, Lobanov2003}; see also \citealt{LobanovZensus2001}). These regimes may correspond to the oscillations of the instantaneous opening angle detected in the M87 jet, at $z\lesssim 1000\; r_{\rm g}$ and $z\gtrsim 2000\; r_{\rm g}$ respectively (\citealt{Mertens2016}, their Figure 6). In this interpretation, the intermediate case (i.e. $1000\; r_{\rm g}\lesssim z\lesssim 2000\; r_{\rm g}$, where oscillations are not clear), would correspond to the transition from a Poynting to a kinetic-energy dominated jet, driven by the kink instability in its non-linear regime.

\section{Conclusions}
\label{sec:conclusions}

We have explored the effect of the kink instability on cylindrical, force-free jets in the linear regime. To get robust conclusions, we have considered a large class of background solutions \citep{Mizuno2012}. In principle, the growth rate can depend on all the parameters describing the background fields, namely: (i) the angular velocity, $\Omega_0$; (ii) its slope at large radii, $\beta$; (iii) the logarithmic derivative (calculated at the core scale) of the poloidal magnetic field, $\alpha$.

Calculating the dispersion relation for different combinations of the parameters, we have found that the group velocity of the perturbation, $v_{\rm p}$, is closely related to the velocity of the plasma in the background fields. Hence, one can estimate
\begin{equation}
\label{eq:concl_2}
\gamma_{\rm p}\approx\Gamma\;,
\end{equation}
where $\Gamma$ is the typical Lorentz factor of the plasma and $\gamma_{\rm p}\equiv~1/\sqrt{1-v_{\rm p}^2/c^2}$. This is a natural result since the instability develops in the plasma comoving frame.

The growth rate of the instability (corresponding to the most unstable wavelength) can be expressed in terms of $\alpha$ and $\gamma_{\rm p}$, while it is insensitive to $\beta$. We have provided a simple equation reproducing our results:
\begin{equation}
\label{eq:concl_1}
\omega_{\rm i,max}\approx 0.17\times\frac{\alpha}{\gamma_{\rm p}}\; \frac{c}{r_0} \;,
\end{equation}
where $r_0$ is the scale of the core (note that narrow jets are more unstable).
In particular, the growth rate is suppressed due to time dilation from the rest frame of the perturbation (see the factor $1/\gamma_{\rm p}$ in the equation above). We also confirm previous results \citep{Lyubarski1999, Mizuno2012}, finding that the growth rate of the perturbation is severely suppressed for a nearly flat poloidal field (i.e. $\alpha\sim 0$).

Applying these results to the well resolved jet of the active galaxy M87 \citep{Mertens2016}, we have shown that the kink instability becomes non-linear at a distance from the central black hole comparable to where the jet stops accelerating.
This scenario is broadly consistent with both (i) the oscillations of the instantaneous opening angle of the jet, which may be due to the kink/Kelvin-Helmholtz modes, in the Poynting/kinetic-energy dominated regimes respectively; (ii) the variability of the light curve (at all wavelengths shorter than the radio, and in particular at TeV energies; \citealt{Aharonian2006}), suggesting that the size of the region where bright emission is expected due to dissipation is comparable to the transverse scale of the jet at the relevant distance from the source (i.e. where the kink modes become non-linear).
Hence (at least for this object), we have suggested that the kink instability of the jet may be the mechanism driving the transition from a Poynting-dominated to a kinetic-energy-dominated flow.

\section*{Acknowledgements}

ES and YEL acknowledge support from the Israeli Science Foundation under Grant No. 719/14.
MCS acknowledges support from the Deutsche Forschungsgemeinschaft in the Collaborative Research Center (SFB 881) ``The Milky Way System'' (subprojects B1, B2, and B8) and in the Priority Program SPP 1573 ``Physics of the Interstellar Medium'' (grant numbers KL 1358/18.1, KL 1358/19.2). MCS furthermore thanks the European Research Council for funding in the ERC Advanced Grant STARLIGHT (project number 339177).

\def\aap{A\&A}\def\aj{AJ}\def\apj{ApJ}\def\apjl{ApJ}\def\mnras{MNRAS}
\def\araa{ARA\&A}\def\physrep{PhR}\def\sovast{Sov. Astron.}\def\nar{NewAR}
\def\aapr{Astronomy \& Astrophysics Review}\def\apjs{ApJS}\def\nat{Nature}\def\na{New Astron.}
\def\prd{Phys. Rev. D}\def\pre{Phys. Rev. E}\def\pasp{PASP}
\bibliographystyle{mn2e}
\bibliography{2d}

\appendix

\section{Instability of the high-\MakeLowercase{m} modes}
\label{sec:appendixA}

\begin{figure}{\vspace{3mm}} 
\centering
\includegraphics[width=0.49\textwidth]{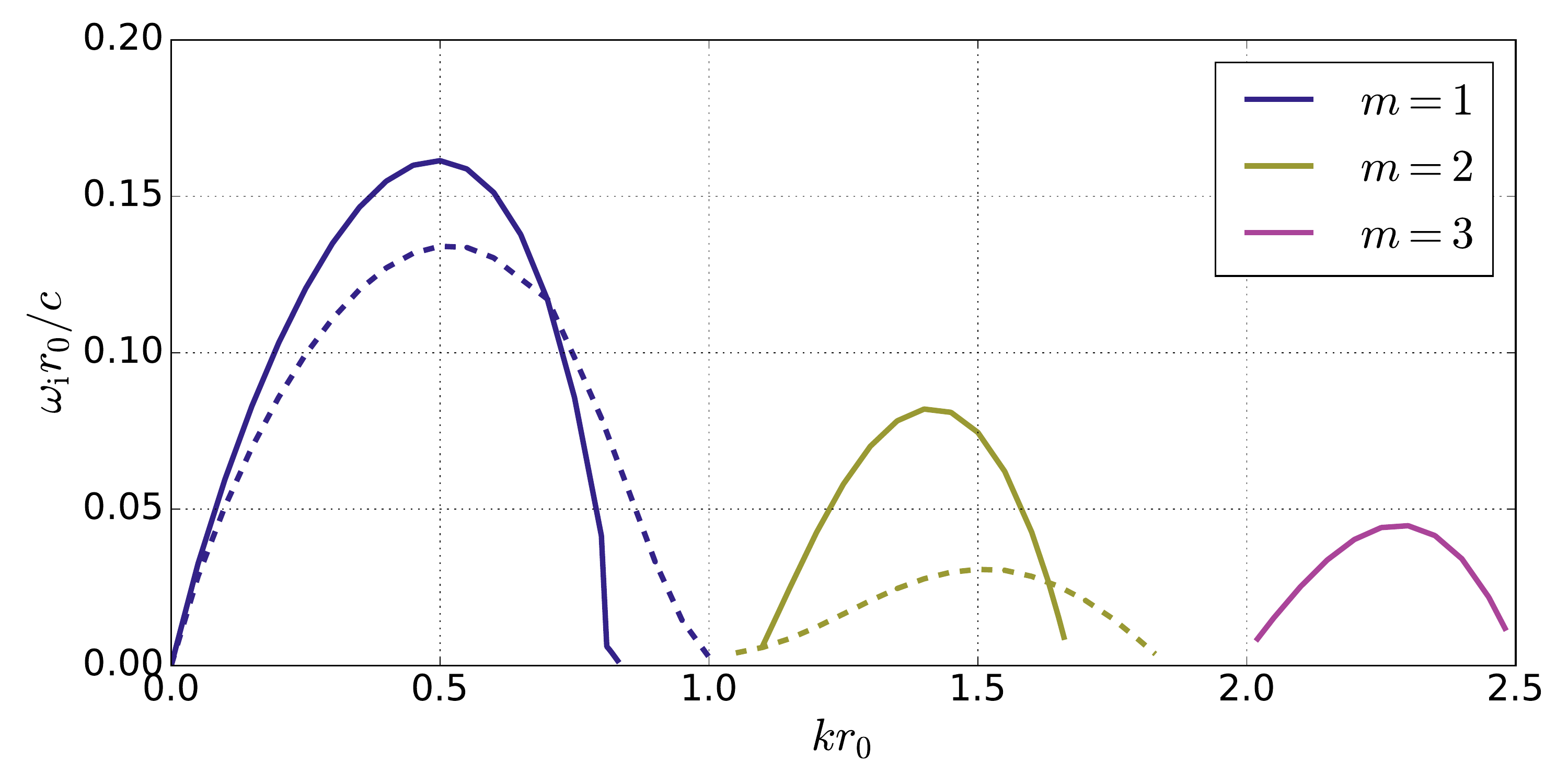}
\caption{Dispersion relation for $\omega_{\rm i}$ in a force-free jet. Different peaks correspond to modes with different $m$, while solid/dashed curves show $\Omega_0r_0/c=0.0/1.0$. The magnetic field profile and the asymptotic slope of the angular velocity are defined by $\alpha=0.7$ and $\beta=1.0$ respectively.}
\label{fig:dispersion_m}
\end{figure}

When $m>1$, solving Eq. \eqref{eq:main} one needs to pay attention to the fact that $G\sim r^3$, but $D\sim r$ when $r\sim 0$. Hence, we have to take the proper initial conditions, i.e. $f\left(0\right)=0$ and $f'\left(0\right)=1$ (here the value of the derivative corresponds to an arbitrary normalisation).

In Figure \ref{fig:dispersion_m} we show the dispersion relation for $\omega_{\rm i}$, where different peaks correspond to modes with different $m$. At high $m$, the peak shifts to large $k$ and the instability is suppressed as expected. Interestingly, this suppression is stronger for the dashed ($\Omega_0 r_0/c=1.0$) than for the solid curves ($\Omega_0 r_0/c=0.0$); hence, at least in the linear regime, the high-$m$ modes may be even less important in the relativistic case. In particular, we were unable to find the dispersion relation when $m=3$ and $\Omega_0 r_0/c=1.0$, probably because of numerical issues when the jet becomes almost stable (i.e. $\omega_{\rm i}r_0/c\ll 1$).

\end{document}